\documentclass[floatfix,aps,superscriptaddress,prd,amsmath,nofootinbib,preprintnumbers,twocolumn]{revtex4}

\usepackage{graphicx}
\usepackage{bm}
\usepackage{float}
\usepackage[
colorlinks=true,        % color link
citecolor=blue,         % cite color
linkcolor=blue,         % link color
urlcolor=blue           % url color
]{hyperref}             % create hyperlinks

\newcommand{\nc}{\newcommand*}
\nc{\Om}{\Omega}
\nc{\ogw}{\Omega_{\mathrm{GW}}}
\nc{\rd}{\mathrm{d}}
\nc{\eg}{\textit{e.g.~}}
\nc{\red}[1]{\textcolor{red}{#1}}
\nc{\lvc}{LIGO/Virgo} % LIGO-VIRGO collaboration

\def\({\left(}
\def\){\right)}
\def\[{\left[}
\def\]{\right]}
\def\e{\begin{equation}}
\def\q{\end{equation}}
\def\m{\begin{eqnarray}}
\def\n{\end{eqnarray}}

\begin{document}

\title{Probing the stochastic signal from primordial gravitational waves with pulsar timing arrays}

%%%%%%%%%%%%%%%%%%%%%%%%%%%%%%%%%%%% author %%%%%%%%%%%%%%%%%%%%%%%%%%%%%%%%%%%%
\author{Jun~Li}
%\author{Jun~Li\orcidlink{0000-0001-5173-271X}}
\email{lijun@qust.edu.cn}
\affiliation{Qingdao Key Laboratory of Novel Optoelectronic Devices and Ultrafast Intelligent Manufacturing, School of Mathematics and Physics, Qingdao University of Science and Technology, Qingdao 266061, China.}
\affiliation{CAS Key Laboratory of Theoretical Physics, Institute of Theoretical Physics, Chinese Academy of Sciences, Beijing 100190, China}

%%%%%%%%%%%%%%%%%%%%%%%%%%%%%%%%%%%% author %%%%%%%%%%%%%%%%%%%%%%%%%%%%%%%%%%%%
\author{Guanghai Guo}
\affiliation{Qingdao Key Laboratory of Novel Optoelectronic Devices and Ultrafast Intelligent Manufacturing, School of Mathematics and Physics, Qingdao University of Science and Technology, Qingdao 266061, China.}

%%%%%%%%%%%%%%%%%%%%%%%%%%%%%%%%%%%% author %%%%%%%%%%%%%%%%%%%%%%%%%%%%%%%%%%%%
\author{Pengfei Yan}
\affiliation{Qingdao Key Laboratory of Novel Optoelectronic Devices and Ultrafast Intelligent Manufacturing, School of Mathematics and Physics, Qingdao University of Science and Technology, Qingdao 266061, China.}

\date{\today}

\begin{abstract}
In this study, we investigate the scenario in which the stochastic signal arises from primordial gravitational waves. Within this framework, we consider two distinct possibilities: one in which the pulsar timing arrays (PTAs) signal corresponds to a stochastic gravitational-wave background (SGWB), and one in which it does not. Primordial gravitational waves can generate an SGWB spanning an exceptionally broad frequency range and are also a source of B‑mode polarization in the cosmic microwave background (CMB). We combine CMB B-mode polarization data from BICEP/Keck (BK18), Planck (Planck18), and baryon acoustic oscillation (BAO) measurements with SGWB limits from PTAs to derive updated constraints on the tensor spectral index of the primordial power spectrum. Under the assumption of no detection of an SGWB from PTAs, the allowed parameter space excludes a large portion of the positive region. The constraint within PTA limits is $n_t= -0.165^{+1.20}_{-1.56}$ at $95\%$ confidence level, which are consistent with those obtained from the combined BK18+Planck18+BAO dataset, leading to tighter constraints on the tensor spectral index. Conversely, if the PTA signal is interpreted as an SGWB, the likelihood distribution for the tensor spectral index favors positive values, with $n_t= 2.39^{+1.46}_{-1.35}$ at $95\%$ confidence level, providing evidence for a blue-tilted primordial gravitational-wave power spectrum. In this case, the allowed parameter space excludes the negative region.
\end{abstract}

\maketitle

%%%%%%%%%%%%%%%%%%%%%%%%%%%%%%%%%%%%%%%%
%%%%%%%%%%%%%%%%%%%%%%%%%%%%%%%%%%%%%%%
\section{introduction}
The observation of gravitational waves (GWs) from coalescing compact binaries with ground‑based interferometers \cite{LIGOScientific:2018mvr,LIGOScientific:2020ibl,KAGRA:2021vkt} has reshaped our view of the cosmos and substantially enhanced our capacity to test gravitational theories in the strong‑field regime. The detections have not only verified the existence of GWs but also yielded a rich trove of data on the physical and astrophysical nature of their sources. The detection of individual GW events is a milestone achievement. However, the observation of another class of source, the stochastic gravitational-wave background (SGWB), remains an ongoing endeavor. Its detection and characterization are of profound importance, promising to illuminate a range of cosmological processes in the early Universe and diverse astrophysical phenomena.

Pulsar timing arrays (PTAs) have become an indispensable observational instrument to probe the SGWB in the nanohertz frequency regime. The nanohertz band accessible to PTAs coincides with the characteristic frequency range of GWs produced by a variety of cosmological and astrophysical processes, positioning PTAs as a premier tool for uncovering SGWBs that may originate from the early Universe or fundamental physics beyond the Standard Model. Recently, the PTA community has achieved substantial progress. The observational results reported by the Chinese PTA (CPTA) \cite{Xu:2023wog}, the European PTA (EPTA) along with the Indian PTA (InPTA) \cite{EPTA:2023fyk,EPTA:2023sfo,EPTA:2023xxk}, the Parkes PTA (PPTA) \cite{Reardon:2023gzh,Zic:2023gta,Reardon:2023zen}, and the North American Nanoherz Observatory for GWs (NANOGrav) \cite{NANOGrav:2023gor,NANOGrav:2023hde} collaborations could be explained as an SGWB. Simultaneously, the inferred amplitude and spectrum of the PTA signal are consistent with astrophysical predictions for a signal originating from the population of supermassive black hole binaries.

The nanohertz frequency band spans a rich variety of cosmological phenomena capable of generating SGWB. With current data, it remains challenging to rule out a cosmological origin for the observed signal. Among the most extensively studied scenarios capable of producing an SGWB in the nanohertz band are scalar-induced gravitational waves generated from enhanced primordial fluctuations \cite{Lopez:2025gfu,Cecchini:2025oks,Appleby:2023igj,Bernardo:2023bqx,Bernardo:2023mxc,Tagliazucchi:2023dai,Balaji:2023ehk,Domenech:2019quo,Domenech:2021ztg,Choudhury:2023hvf,Choudhury:2023rks,Choudhury:2023jlt,Choudhury:2013woa,Fu:2019vqc,Pi:2020otn,
Hajkarim:2019nbx,Domenech:2020kqm,Yi:2020kmq,Tomikawa:2019tvi,Inomata:2019yww,Hwang:2017oxa,Domenech:2020xin,Yuan:2020iwf,Yuan:2019udt,Jinno:2013xqa,Chen:2019xse,Bartolo:2018rku,Tada:2019amh,Espinosa:2018eve,Di:2017ndc,
Jin:2023wri,You:2023rmn,Orlofsky:2016vbd,Alabidi:2013lya,Osano:2006ew,Noh:2004bc,Matarrese:1997ay,Giovannini:2010tk,Xu:2019bdp,Unal:2018yaa,Cai:2018dig,Cai:2019jah,Cai:2019elf,Cai:2019amo,Assadullahi:2009nf,Assadullahi:2009jc,
Inomata:2019zqy,Inomata:2019ivs,Inomata:2018epa,Yuan:2019fwv,Zhou:2020kkf,Alabidi:2012ex,Kohri:2018awv,Lu:2019sti,Li:2022avp,Ananda:2006af,Baumann:2007zm,Inomata:2016rbd,Li:2021uvn,Chen:2024fir,Li:2023uhu}, primordial gravitational waves \cite{Ezquiaga:2021ler,Saikawa:2018rcs,Campeti:2020xwn,Cai:2016ldn,Giare:2020vss,Brax:2017pzt,Dubovsky:2009xk,Lin:2016gve,Cai:2020ovp,Li:2017cds,Li:2018iwg,Li:2019efi,Li:2019vlb,Li:2021scb,Li:2021nqa,Li:2024cmk},   and many others. 

In this study, we investigate the scenario in which the stochastic signal arises from primordial gravitational waves. Within this framework, we consider two distinct possibilities: either the signal detected by PTAs constitutes an SGWB, or it does not. Primordial gravitational waves can generate an SGWB that spans an exceptionally broad range of frequencies, and are also a source of B-mode polarization in the cosmic microwave background (CMB). We combine CMB B-mode polarization data from the BICEP/Keck observations \cite{BICEP:2021xfz}, Planck observations \cite{Planck:2018vyg} and Baryon Acoustic Oscillation (BAO) measurements \cite{Beutler:2011hx, Ross:2014qpa, BOSS:2016wmc} including the latest DESI Data Release 2 results \cite{DESI:2025zgx}, along with SGWB limits from PTAs to derive updated constraints on the tensor spectral index of the primordial power spectrum.

\section{stochastic signal from primordial gravitational waves}
A stochastic gravitational-wave background arises from the incoherent superposition of a vast number of weak, independent, and unresolved gravitational-wave sources. To characterize its spectral properties, it is useful to describe the distribution of energy across frequency, typically expressed in terms of the energy density spectrum
\e
\Omega_{\mathrm{gw}}(f)=\frac{1}{\rho_c}\frac{d\rho_{\mathrm{gw}}}{d\ln f}=\frac{2\pi^2}{3H_0^2}f^3 S_h(f).
\q
The fractional energy density in gravitational waves is a dimensionless measure of the gravitational-wave contribution to the total energy density of the Universe. Meanwhile, $S_h(f)$ denotes the strain power spectral density. 

A stochastic background of primordial gravitational waves is generated by the superadiabatic amplification of zero-point quantum fluctuations in the gravitational field. The power spectrum of tensor perturbations is commonly parameterized as
\m
P_t(k)&=&A_t\(\frac{k}{k_*}\)^{n_t},\label{eqs:spectrumtensor}
\n
where $A_t$ is the tensor amplitude at the pivot scale $k_*=0.01$ Mpc$^{-1}$ and $n_t$ is the tensor spectral index.
For convenience, we quantify the tensor amplitude relative to the scalar amplitude $A_s$ by introducing the tensor-to-scalar ratio, defined as
\e
r\equiv\frac{A_t}{A_s}.
\q
The current intensity of the stochastic gravitational-wave background originating from primordial gravitational waves is given by \cite{Lasky:2015lej,Caprini:2018mtu}
\m
\Omega_{\mathrm{gw}}(f)\approx1.5\times10^{-16}\Big(\frac{r}{0.032}\Big)\Big(\frac{f}{f_*}\Big)^{n_t},
\n
where $f_*\approx1.54\times10^{-17}$ Hz related to the CMB pivot scale $k_*=0.01$ Mpc$^{-1}$.

\section{data and result}
In this study, we combine PTA data from the second data release of the European PTA (EPTA DR2) \cite{EPTA:2023fyk, EPTA:2023sfo, EPTA:2023xxk} with BK18+Planck18+BAO datasets. Based on the EPTA DR2 dataset, we identify five representative sensitivity points, each characterized by a specific frequency and its corresponding fractional energy density, as listed in Table~\ref{table1}. These points are systematically compared to determine which one delivers the most stringent constraint. We then combine the BK18+Planck18+BAO dataset with (i) no PTA data, (ii) each individual PTA sensitivity point (A–E), and (iii) the full PTA dataset.

\begin{table}[htb]
\newcommand{\tabincell}[2]{\begin{tabular}{@{}#1@{}}#2\end{tabular}}
  \centering
  \begin{tabular}{  c |c| c}
  \hline
  \hline
  Point & \tabincell{c} {frequency $f$} & \tabincell{c}{fractional energy density $\Omega_{\mathrm{gw}}$}\\
  \hline
  A & $f=10^{-9}$   &$\Omega_{\mathrm{gw}}h^2=8*10^{-9}$\\
 % \hline
  B &$f=3*10^{-9}$    &$\Omega_{\mathrm{gw}}h^2=4*10^{-10}$\\
 % \hline
  C & $f=10^{-8}$   &$\Omega_{\mathrm{gw}}h^2=8*10^{-9}$\\
 % \hline
  D &  $f=10^{-7}$   &$\Omega_{\mathrm{gw}}h^2=2*10^{-5}$ \\
 % \hline
  E  & $f=3*10^{-8}$     &$\Omega_{\mathrm{gw}}h^2=2*10^{-7}$ \\
  \hline
  \end{tabular}
  \caption{Five representative sensitivity points from the EPTA DR2 dataset, each corresponding to a specific frequency and associated fractional energy density.}
  \label{table1}
\end{table}

We investigate two mutually exclusive hypotheses: (i) the detected PTA signal originates from a stochastic gravitational-wave background, or (ii) it does not. At a reference frequency of $f=10^{-9}$, the fractional energy density satisfies $\Omega_{\mathrm{gw}}h^2\ge8*10^{-9}$ under the SGWB hypothesis, and $\Omega_{\mathrm{gw}}h^2<8*10^{-9}$ if it does not.

We constrain the tensor spectral index using current observational data through the publicly available Cosmomc package \cite{Lewis:2002ah}. In this analysis, the parameters $r$ and $n_t$ are treated as completely free, with uniform priors $r\in[0, 2]$, and $n_t\in[-10, 10]$. The standard $\Lambda$CDM cosmological parameters are fixed to the Planck 2018 best-fit values \cite{Planck:2018vyg}: $\Omega_bh^2=0.02242$, $\Omega_ch^2=0.11933$, $100\theta_{\mathrm{MC}}=1.04101$, $\tau=0.0561$, $\ln\(10^{10}A_s\)=3.047$, and $n_s=0.9665$.

Within the $\Lambda$CDM+$r$+$n_t$ model, the constraints from the BK18+Planck18+BAO datasets alone are:
\m
r &<& 0.041\quad(95\% \ \mathrm{C.L.}),\\
n_t &=&  1.42^{+2.34}_{-2.63}\quad(95\%\ \mathrm{C.L.}).
\n

\begin{figure}
    \centering
    \includegraphics[width=8.8cm]{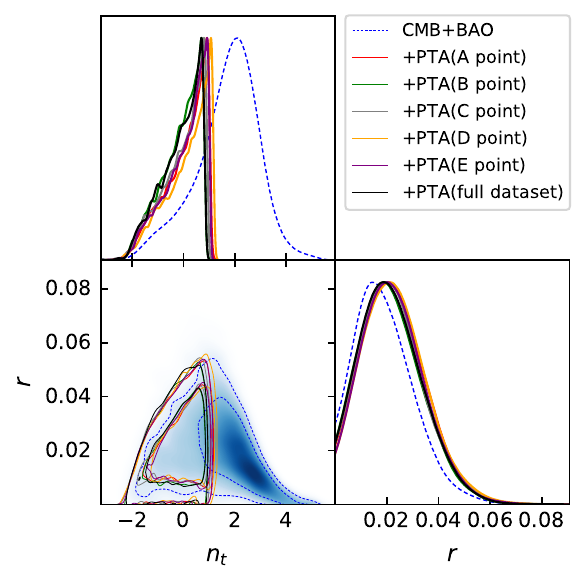}
    \caption{The marginalized posterior contours and likelihood distributions for the tensor spectral index $n_t$ and the tensor-to-scalar ratio $r$ are shown at  $68\%$ and $95\%$ confidence levels, derived from the BK18+Planck18+BAO dataset combined respectively with: (i) no PTA data, (ii) PTA sensitivity point A, (iii) point B, (iv) point C, (v) point D, (vi) point E, and (vii) the full PTA dataset. In all cases, we assume no detection of an SGWB from the PTA observations.}
 \label{fig1}
\end{figure}

Assuming no detection of an SGWB from PTA, we combine BK18+Planck18+BAO with individual PTA sensitivity points (A–E) and with the full PTA dataset. The resulting constraints on $r$ and $n_t$ are summarized below:\\
$\cdot$ BK18+Planck18+BAO+PTA (A point)
\m
r &<& 0.044\quad(95\% \ \mathrm{C.L.}),\\
n_t &=& 0.024^{+1.25}_{-1.68}\quad(95\%\ \mathrm{C.L.});
\n
$\cdot$ BK18+Planck18+BAO+PTA (B point)
\m
r &<& 0.043\quad(95\% \ \mathrm{C.L.}),\\
n_t &=& -0.156^{+1.18}_{-1.55}\quad(95\%\ \mathrm{C.L.});
\n
$\cdot$ BK18+Planck18+BAO+PTA (C point)
\m
r &<& 0.044\quad(95\% \ \mathrm{C.L.}),\\
n_t &=& -0.083^{+1.23}_{-1.65}\quad(95\%\ \mathrm{C.L.});
\n
$\cdot$ BK18+Planck18+BAO+PTA (D point)
\m
r &<& 0.045\quad(95\% \ \mathrm{C.L.}),\\
n_t &=& 0.140^{+1.28}_{-1.72}\quad(95\%\ \mathrm{C.L.});
\n
$\cdot$ BK18+Planck18+BAO+PTA (E point)
\m
r &<& 0.044\quad(95\% \ \mathrm{C.L.}),\\
n_t &=& 0.028^{+1.22}_{-1.67}\quad(95\%\ \mathrm{C.L.});
\n
$\cdot$ BK18+Planck18+BAO+PTA (full dataset)
\m
r &<& 0.044\quad(95\% \ \mathrm{C.L.}),\\
n_t &=& -0.165^{+1.20}_{-1.56}\quad(95\%\ \mathrm{C.L.}).
\n
These results are presented graphically in Fig.~\ref{fig1}. In the absence of PTA data, the likelihood distribution for the tensor spectral index $n_t$ shows a preference for positive values. Under the assumption of no detection of an SGWB from PTAs, the allowed parameter space excludes a large portion of the positive region. The constraints from PTA observations are consistent with those obtained from the combined BK18+Planck18+BAO datasets, leading to a further tightening of the constraints on the tensor spectral index. A scale-invariant primordial gravitational-wave power spectrum remains compatible with current data, and the inclusion of PTA measurements significantly strengthens the constraints on the positive side of the tensor spectral index.

\begin{figure}[thp]
    \centering
    \includegraphics[width=7.5cm]{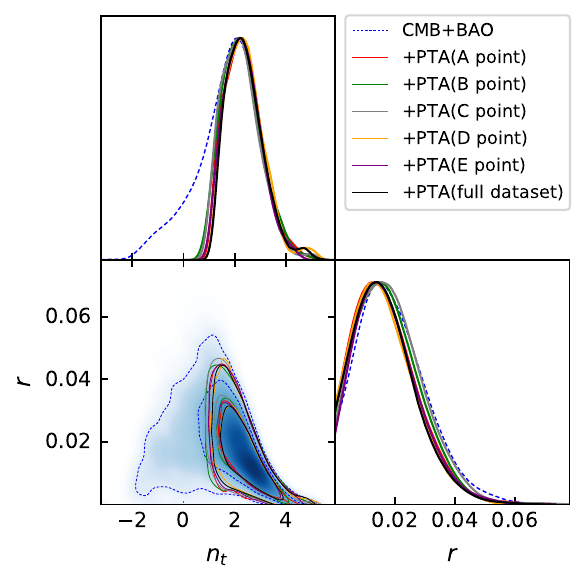}
    \caption{The marginalized posterior contours and likelihood distributions for the tensor spectral index $n_t$ and the tensor-to-scalar ratio $r$ are shown at  $68\%$ and $95\%$ confidence levels, derived from the BK18+Planck18+BAO dataset combined respectively with: (i) no PTA data, (ii) PTA sensitivity point A, (iii) point B, (iv) point C, (v) point D, (vi) point E, and (vii) the full PTA dataset. In this analysis, the PTA signal is assumed to correspond to a detection of SGWB.}
    \label{fig2}
\end{figure}

Alternatively, under the assumption that the PTA signal corresponds to an SGWB detection, we combine BK18+Planck18+BAO with each PTA sensitivity point (A–E) and with the full PTA dataset. The resulting constraints on $r$ and $n_t$ are as follows:\\
$\cdot$ BK18+Planck18+BAO+PTA (A point)
\m
r &<& 0.037\quad(95\% \ \mathrm{C.L.}),\\
n_t &=& 2.34^{+1.49}_{-1.40}\quad(95\%\ \mathrm{C.L.});
\n
$\cdot$ BK18+Planck18+BAO+PTA (B point)
\m
r &<& 0.036\quad(95\% \ \mathrm{C.L.}),\\
n_t &=& 2.26^{+1.59}_{-1.45}\quad(95\%\ \mathrm{C.L.});
\n
$\cdot$ BK18+Planck18+BAO+PTA (C point)
\m
r &<& 0.038\quad(95\% \ \mathrm{C.L.}),\\
n_t &=&2.22^{+1.49}_{-1.40}\quad(95\%\ \mathrm{C.L.});
\n
$\cdot$ BK18+Planck18+BAO+PTA (D point)
\m
r &<& 0.035\quad(95\% \ \mathrm{C.L.}),\\
n_t &=& 2.41^{+1.44}_{-1.39}\quad(95\%\ \mathrm{C.L.});
\n
$\cdot$ BK18+Planck18+BAO+PTA (E point)
\m
r &<& 0.037\quad(95\% \ \mathrm{C.L.}),\\
n_t &=& 2.30^{+1.39}_{-1.36}\quad(95\%\ \mathrm{C.L.});
\n
$\cdot$ BK18+Planck18+BAO+PTA (full dataset)
\m
r &<& 0.036\quad(95\% \ \mathrm{C.L.}),\\
n_t &=& 2.39^{+1.46}_{-1.35}\quad(95\%\ \mathrm{C.L.}).
\n
These numerical results are summarized visually in Fig.~\ref{fig2}. If the PTA signal is interpreted as an SGWB, the likelihood distribution for the tensor spectral index $n_t$ favors positive values, supporting a blue-tilted spectrum for primordial gravitational waves. In this case, the allowed parameter space excludes the negative region.

\section{summary}
In this study, we investigate the scenario in which the stochastic signal arises from primordial gravitational waves. Within this framework, we consider two mutually exclusive possibilities: either the signal detected by PTAs constitutes an SGWB, or it does not. By combining BK18+Planck18+BAO with the limits on the SGWB from PTAs, we derive updated constraints on the tensor spectral index of the primordial power spectrum. In the absence of PTA data, the likelihood distribution for the tensor spectral index shows a preference for positive values. Under the assumption of no detection of an SGWB from PTAs, the allowed parameter space excludes a large portion of the positive region. Conversely, if the PTA signal is interpreted as an SGWB, the likelihood distribution for the tensor spectral index favors positive values, supporting a blue-tilted spectrum for primordial gravitational waves.

\noindent {\bf Acknowledgments}.
This work is supported by the National Natural Science Foundation of China (Grant No. 12405069), the Natural Science Foundation of Shandong Province (Grant No. ZR2021QA073) and Research Start-up Fund of QUST (Grant No. 1203043003587).

%%%%%%%%%%%%%%%%%%%%%%%%%%%%%%%%%%%%%%%%
%%%%%%%%%%%%%%%%%%%%%%%%%%%%%%%%%%%%%%%%

%%%%%%%%%%%%%%%%%%%%%%%%%%%%%%%%%%%%%%%%
%%%%%%%%%%%%%%%%%%%%%%%%%%%%%%%%%%%%%%%%
%%%%%%%%%%%%%%%%%%%%%%%%%%%%%%%%%%%%%%%%
%%%%%%%%%%%%%%%%%%%%%%%%%%%%%%%%%%%%%%%%
\end{document}